\documentclass{article}
\usepackage[utf8]{inputenc}

\usepackage{amsmath,amsthm,amssymb}
\usepackage[dvipsnames]{xcolor}
\usepackage{physics}
\usepackage{tensor}
\usepackage{graphicx}
\usepackage{float}
\usepackage{xcolor, colortbl}
\usepackage{hyperref}
\usepackage{authblk}
\usepackage{mathtools}

\definecolor{light-gray}{gray}{0.9} 

\usepackage{setspace}
\usepackage{xparse}
\usepackage{esvect} %for vector notation
\usepackage{slashed} %for Feynman slash notation
\usepackage{dsfont} %for the real numbers symbol R
\usepackage[dvipsnames]{xcolor} %for textcolor
\usepackage{empheq}
\usepackage{makecell}
\usepackage{subcaption}
\newenvironment{eqaed}
    {
    \begin{equation}
        \begin{aligned}
    }
    {
        \end{aligned}
    \end{equation}
    }
\usepackage{float}
\usepackage{hyperref}
\hypersetup{
    colorlinks=true,
    linkcolor=blue,
    filecolor=magenta,      
    urlcolor=cyan,
}
\usepackage{import}
\usepackage{makeidx}
\title{On Higher Dimensional Self-Similar \\
Axion-Dilaton Solutions}\author[1]{Ehsan Hatefi}
\author[2]{Eleonora Vanzan\footnote{ehsan.hatefi@sns.it, eleonora.vanzan@studenti.unipd.it}}
\affil[1]{Scuola Normale Superiore and I.N.F.N,\protect\\ 
Piazza dei Cavalieri 7, 56126, Pisa, Italy\vspace{.7em}}
\affil[2]{Dipartimento di Fisica e Astronomia “Galileo Galilei”,\protect\\ Universit\`a di Padova, 35131 Padova, Italy}

\begin{document}
\maketitle
\vspace{-0.9cm}
\begin{abstract}
    \noindent
    
    We show that solutions of the self-similar gravitational collapse in the Einstein-axion-dilaton system exist in higher--dimensional spacetimes. These solutions are invariant under spacetime dilation combined with internal SL(2,$\mathds{R}$) transformations. We rely on the recent setup of \cite{Antonelli:2019dqv} and use it for the three different conjugacy classes (elliptic, parabolic and hyperbolic) in higher dimensions. Lastly, we identify new families of physically distinguishable self-similar solutions for all three conjugacy classes in six and seven dimensions.
\end{abstract}

\newpage

\section{Introduction}

An interesting thought experiment for critical phenomena in gravitation was proposed by M. Choptuik in \cite{Chop} (see \cite{Gundlach} for more references). Choptuik studied the spherical gravitational collapse of a scalar field, distinguishing the initial conditions that lead to collapse solutions with black hole formation from those that lead to the empty Minkowski space \cite{Chop,AlvarezGaume:2008fx}. 

Generally speaking, one can consider a one--parameter class of initial conditions labelled by the value of the initial field amplitude $p$. If $p$ is small, the time evolution is linear and the collapse does not take place. Instead, for large $p$ a so--called trapped surface forms and a black hole should exist in the final state. Therefore, there is a value $p_\text{crit}$ that marks the transition between these two regimes. A scaling law can be illustrated, and for supercritical initial conditions $p>p_\text{crit}$ one finds that the mass of the black hole scales as
\begin{equation}
M_{\rm bh}(p) \propto (p-p_\text{crit})^\gamma\,,
\end{equation}
with the Choptuik exponent $\gamma\approx0.37$ (see also \cite{Gundlach,Hamade:1995ce}).
The  solutions on the so called ``critical surface'' realize some form of spacetime self-similarity. This phenomenon has led to various research topics that can be potentially related to critical phenomena, as well as to the issue of scale invariance in gravity.
For a real scalar field, the solutions show a discretely self-similar behaviour, which is difficult to deal with algebraically. On the other hand, Continuous Self-Similarity (CSS) means that there is an invariance under continuous one-parameter groups of homotheties, which reproduces itself once one introduces matter fields with internal symmetries. Therefore, one can look for solutions that are invariant under some combinations of scalings and internal symmetry transformations.

Here we focus on the axion-dilaton system, which does experience gravitational collapse with the fascinating property of Choptuik scaling.
However, various numerical results with different matter fields have been carried out. For example, critical solutions for a massless scalar field were obtained in \cite{Sorkin_2005,Bland_2005,Birukou:2002kk,Husain:2002nk}, and an interesting work on Einstein-Maxwell-dilaton theories and critical collapse was recently done in \cite{Rocha:2018lmv}. The specific case of a complex scalar field was studied in \cite{Hirschmann:1994du}. The authors in \cite{AlvarezGaume:2008qs,evanscoleman,KHA,MA} dealt with the critical collapse of the radiation fluid, while \cite{Hirschmann_1997} considered non-linear $\sigma$-model computations. Interestingly, \cite{Hirschmann_1997} has some overlap with our paper, because it scrutinizes the elliptic solutions of the axion-dilaton configuration. The scaling in vacuum axisymmetric gravitational collapse was considered in \cite{AE}.
The correspondence between critical collapse and self-similarity was thus explored beyond the well known spherically symmetric real scalar fields. The axion-dilaton system was studied in the past only in four dimensions \cite{Hirschmann:1994du,Hamade:1995jx,Eardley:1995ns} and also recently done in four and five dimensions \cite{Antonelli:2019dqv}. The aim of the present paper is not only to confirm that there are indeed self-similar solutions for gravitational collapse in higher dimensions, but also to show that one can identify entire families of physically distinguishable self-similar solutions.

The content of this paper is motivated by the AdS/CFT correspondence \cite{Maldacena:1997re}, by some holographic description of black hole formation \cite{scalingqcd}, and by the physics of black holes and its applications \cite{Hatefi:2012bp}.
In type IIB String Theory and AdS/CFT, one may investigate the gravitational collapse on spaces that approach asymptotically to $AdS_5 \times S^5$, and a natural choice for the matter content involves the axion-dilaton and the self-dual 5-form field.
It can be shown that in five dimensions the simplest dynamical setting is the Einstein-axion-dilaton system with a cosmological constant. However, Einstein spaces do not admit homothetic vector fields, which could be an apparent problem when considering self-similar collapse. Nevertheless, we are dealing with critical gravitational collapse and we are only interested in a small spacetime region, close to the place where the singularity happens. Hence, we are no longer concerned about the asymptotic structure of the spacetime (this was supported by numerical evidence, as seen in \cite{Birukou:2002kk,Husain:2002nk}). Henceforth, we shall drop the cosmological constant and analyse self-similar critical collapse in dimensions ranging from four up to seven for all elliptic, parabolic and hyperbolic cases accordingly.

%\textcolor{blue}{}

The organisation of this paper is as follows. First, we briefly describe the axion-dilaton system and the so called CSS ansatz. For the sake of completeness, we write the equations of motion and discuss the initial conditions for all three conjugacy classes of internal SL(2,$\mathds{R}$) transformations that must be applied to compensate the scaling transformations in spacetime. We then shortly highlight self-similar solutions for all three classes in four and five dimensions, which provide the necessary backgrounds for higher--dimensional spacetimes.
Lastly, we discover new solutions in six and seven dimensions for all three conjugacy classes, which entails a number of distinguished CSS solutions for six different cases. For instance, in six and seven dimensions for the elliptic class, which have $U(1)$ compensated dilations, we shall explore \emph{four} and \emph{three} distinct collapse solutions.

\section{Axion-dilaton and continuous self-similarity}

One can combine two real scalars, the axion $a$ and dilaton $\phi$, into a complex scalar field $\tau \equiv a + i e^{- \phi}$, and its coupling to gravity in $d\geq 4$ dimensions is determined by the action
\begin{equation}
S = \int d^d x \sqrt{-g} \left( R - {1 \over 2} { \partial_a \tau
\partial^a \bar{\tau} \over (\mathop{\rm Im}\tau)^2} \right) \; .
\label{eaction}\end{equation}
 where $R$ is the scalar curvature. The corresponding equations of motion read
\begin{eqnarray}
\label{eoms}
R_{ab} - {1 \over 4 (\mathop{\rm Im}\tau)^2} ( \partial_a \tau \partial_b
\bar{\tau} + \partial_a \bar{\tau} \partial_b \tau) & = & 0 \ , \end{eqnarray}
\begin{eqnarray}
\label{eoms14}
\nabla^a \nabla_a \tau + { i \nabla^a \tau \nabla_a \tau \over
\mathop{\rm Im}\tau} = 0 \ .
\end{eqnarray}
The effective action of the axion-dilaton system is classically invariant under  SL(2,$\mathds{R}$) transformations 
\begin{equation}
\tau \rightarrow {a\tau+b \over c\tau+d} \; ,
\label{sltwo}
\end{equation}
where $(a,b,c,d) \in \mathds{R}$, $ad - bc = 1$. The SL(2,$\mathds{R}$) symmetry is broken to an $\text{SL}(2,\mathbb{Z})$ subgroup once one takes into consideration the non-perturbative phenomena (more details can be found out in \cite{SEN_1994,Schwarz_1995,Font:1990gx, greenschwarzwitten, polchinski}).

We assume spherical symmetry and Continuous Self-Similarity (CSS) and look for solutions, as was discussed in \cite{Hamade:1995jx,Eardley:1995ns}, with a metric of the form 
\begin{equation}
	ds^2 = \left(1+u(t,r)\right)\left(- b(t,r)^2dt^2 + dr^2\right)
			+ r^2d\Omega^2_{q} \; ,
\label{metric1}
\end{equation}
where $q=d-2$. CSS implies the existence of a homothetic Killing vector, denoted by $\xi$ in the following, which generates the so called global scale transformation 
\begin{equation}
 {\cal L}_{\xi} g_{ab} = 2 g_{ab}\ .
\label{tauxi}
\end{equation}
Having introduced the scale invariant variable $z=- r/t$, CSS means that the two functions $u(t,r)$ and $b(t,r)$ in the metric only depend upon $z$, so that $b(t,r) = b(z)$, $u(t,r) = u(z)$. On the other hand $\tau(t,r)$ can be invariant up to an SL(2,$\mathds{R}$) transformation, and in  \cite{AlvarezGaume:2011rk} we described in detail the CSS condition for $\tau(t,r)$. Making use of an SL(2,$\mathds{R}$) transformation, one can indeed compensate the action of the homothetic vector field
\begin{equation}
\xi = t \frac{\partial}{\partial t} + r\frac{\partial}{\partial r} \ .
\end{equation}

There are actually three different possible ansatz\"e for the axion-dilaton configuration, depending on whether the SL(2,$\mathds{R}$) transformation used to compensate the scaling transformation in spacetime belongs to the elliptic, parabolic or hyperbolic class.
For all three distinct cases, $\omega$ is a real constant to be found demanding that the critical solution be regular, and $f(z)$ is an arbitrary complex function.

The {\bf elliptic ansatz} is
\begin{equation}
 \tau(t,r)	=  i { 1 - (-t)^{i \omega} f(z) \over 1 + (-t)^{i
\omega} f(z)} ,
\label{tauansatz}
\end{equation}
and in this case a scaling transformation $t\rightarrow \lambda\, t$, $\tau(t,r)$ is accompanied by an SL(2,$\mathds{R}$) rotation.
 
The condition on $f(z)$ for the elliptic class is $\abs{f(z)} < 1$, and 
the infinitesimal form of this type of tranformation reads
\begin{equation}
    \mathcal{L}_\xi \tau = \frac{\omega}{2} \left( \tau^2 +1\right)\,.  
\end{equation}

The {\bf parabolic ansatz} is
\begin{equation}
\tau(t,r) = f(z)+\omega \log(-t) \ ,
\label{tau3ansatz}
\end{equation}
where a scaling transformation is compensated by
a translation, and $f(z)$ is an arbitrary complex function, subject to the only condition  $\Im f(z)>0$. Now the infinitesimal form is
\begin{equation}
    \mathcal{L}_\xi \tau = \omega\,, 
\end{equation}
 and it is worth emphasizing the presence of a new type of transformation that is only relevant to the parabolic case. Indeed, if we perform the scaling
\begin{equation}\label{parabolic_rescaling}
    \omega \rightarrow K \omega\,,\quad f(z) \rightarrow K f(z)\,,\quad K \in \mathbb{R}_+
\end{equation}
$\tau$ also transforms, and
$\tau\rightarrow K \tau$, which is a new kind of symmetry. 

The {\bf hyperbolic ansatz} is
\begin{equation}
\tau(t,r) = \;\frac{1- (-t)^{\omega} f(z)}{1+ (-t)^{\omega} f(z)} \ ,
\label{tau2ansatz}
\end{equation}
where under a scaling transformation $t\rightarrow \lambda\, t$, $\tau(t,r)$ undergoes
an SL(2,$\mathds{R}$) boost, and the condition on $f(z)$ is $\Im f(z)>0$. One can show that using an SL(2,$\mathds{R}$) transformation, the following  ansatz
\begin{equation}
\tau(t,r)= \;\rightarrow (-t)^{\omega}\,f(z).
 \label{tau331ansatz}
\end{equation}
can be chosen for hyperbolic case, and this gives rise to the same e.o.m.'s. In this case the infinitesimal form is
\begin{equation}
    \mathcal{L}_\xi \tau = \omega \tau\,.
\end{equation}
The case $\omega=0$ leads us to the trivial solution $f(z)=\text{constant}$, $b(z)=1$\footnote{These are the initial conditions that produce the flat spacetime with a constant $\tau_0$.}.
\subsection{The equations of motion}
As we have explained in \cite{Antonelli:2019dqv}, taking into account the spherical symmetry one can show that all $u(z)$, $b(z)$ functions can be expressed in terms of $f(z)$. Indeed, the Einstein equations for the angular variables give
\begin{equation}
u(z)\,=\frac{zb'(z)}{(q-1)b(z)} \ ,
\end{equation}
and therefore one can eliminate $u(z)$ from the actual computations and deal with the equations of motion for just $b(z)$ and $f(z)$. Having done some simplifications, one thus arrives at a first-order linear inhomogeneous equation for $b(z)$, expressed just in terms of $f(z)$, $f'(z)$. There is also a second order ordinary differential equation for $f(z)$, whose initial conditions are determined demanding the smoothness of the solution, which also determines the value of $\omega$.

Let us now display the reduced forms of the equations of motion for $f(z)$ and $b(z)$ for the three classes. These equations hold in an arbitrary number of dimensions. Our e.o.m.'s are in agreement with \cite{AlvarezGaume:2011rk}, where they were first derived.
\subsection{Elliptic class}
The e.o.m.'s for self-similar solutions for this particular elliptic class in any dimension $d = q+2$ are
\begin{eqnarray}
b' & = & \frac{ - { 2z(b^2 - z^2)} f' \bar{f}' + {
2i \omega (b^2 - z^2) } (f \bar{f}' - \bar{f} f')
+ {2\omega^2 z |f|^2 }}{q b (1-\abs{f}^2)^2} \,, \nonumber\\
\end{eqnarray}
and the second order ODE for $f(z)$ is
\begin{eqaed}
q z  \left(z^2-b^2\right) (1-\abs{f}^2)^2 f'' = & \,\, b^2 f'  \big( -2 f \left(q z \bar{f}^2 f'-i \omega z \bar{f}'+q^2 \bar{f}\right) \\ 
& \quad -2 z^2 f' \bar{f}'+2 z \bar{f}  (q-i \omega) f'+q^2 \abs{f}^4+q^2\big)\\
& +z  \Big(2 f^2 \left(q (-1-i \omega) z \bar{f}^2 f' +\omega^2 z \bar{f}' -i q \omega \bar{f} \right)\\
&  \quad + f  \left(2 i \omega z^2 f'  \bar{f}' +2 q z^2 \bar{f}^2 f'^2+4 q z \bar{f}  f' +q \omega (\omega+i)\right)\\
&\quad -2 q z f'  \left(z \bar{f}  f' -i \omega+1\right)-q \omega (\omega-i) \abs{f}^2 f\Big)\\
& +\frac{2 z^3}{b^2} \left(z f' -i \omega f \right)^2 \left(z \bar{f}' +i \omega \bar{f} \right) \ .
\end{eqaed}
%Notice that these equations are invariant under a redefinition of the phase of $f(z)$ which means that
The e.o.m.'s are invariant under a residual symmetry of $f(z)$,
\begin{equation}
    f(z) \rightarrow e^{i\theta}f(z)\, .
\end{equation} 
\subsubsection{Parabolic class}
The e.o.m.'s for the parabolic class in any  $d = q+2$  are 
\begin{eqaed}
    b' = -\frac{2 \left(\left(z^2-b^2\right) f' \left(z \bar{f}'-\omega\right)+\omega \left(\left(b^2-z^2\right) \bar{f}'+\omega z\right)\right)}{q b (f-\bar{f})^2} \ ,
\end{eqaed}
\begin{eqaed}
    {q z\left(z^2-b^2\right) (f-\bar{f})^2} f''  = &\,\, b^2 f' \big(2 z f' \left(z \bar{f}'-\omega\right)-2 q f \left(z f'+q \bar{f}\right)\\
    &\quad + 2 q z \bar{f} f'+q^2 f^2-2 \omega z \bar{f}'+q^2 \bar{f}^2\big)\\
    & +z  \Big(2 \omega z \bar{f}' \left(\omega-z f'\right)+2 q f \big(\left(\omega-z f'\right)^2-\bar{f} \left(\omega-2 z f'\right)\big)\\
    &\quad -2 q \bar{f} \left(\omega-z f'\right)^2  +q \bar{f}^2 \left(\omega-2 z f'\right)+q f^2 \left(\omega-2 z f'\right)\Big)\\
    & +\frac{2 z^3}{b^2} \left(\omega-z f'\right)^2 \left(\omega-z \bar{f}'\right)\ .
\end{eqaed}
Note that in this case the e.o.m.'s are invariant under arbitrary shifts of $f(z)$ by a real constant, 
\begin{equation}
    f(z) \rightarrow f(z) + a\, .
\end{equation}
\subsubsection{Hyperbolic class}
Finally, the e.o.m.'s for hyperbolic class in any $d = q+2$ dimension are
\begin{eqaed}
    b' = -\frac{2 \left(\left(z^2-b^2\right) f' \left(z \bar{f}'-\omega \bar{f}\right)+\omega f \left(\left(b^2-z^2\right) \bar{f}'+\omega z \bar{f}\right)\right)}{q b (f-\bar{f})^2}\ ,
\end{eqaed}
\begin{eqaed}
   {q z  \left(z^2-b^2\right) (f-\bar{f})^2} f'' = &\,\, b^2 f' \big(2 z^2 f'\bar{f}'-2 z \omega f \bar{f}'-2\, q\, f (z f'+q \bar{f})\\
   & \quad +2\, q \,z \bar{f} f'+q^2 f^2-2 \omega z \bar{f}{f}'+q^2 \bar{f}^2\big)\\
   +&z \Big(q\omega (1+\omega)f^3-2qz\bar ff'(\bar f-\omega\bar f+zf')\\
   &-2f^2\big(q\omega\bar f+q(1+\omega)zf'-\omega^2z\bar f'\big)\\&
   +f(-q(-1+\omega)\omega\bar f^2+4qz\bar ff'+2z^2f'(qf'-\omega\bar f'))\Big)\\
   &+\frac{2 z^3}{b^2} \left(\omega f-z f'\right)^2(\omega \bar{f}-z \bar{f}') \ .
\end{eqaed}
These equations for the hyperbolic case are invariant under a constant scaling 
\begin{equation}
    f(z) \rightarrow e^\lambda f(z)\,,\quad \lambda \in \mathbb{R}\, .
\end{equation}
\section{Search for solutions and their properties}
For the geometrical point of view, we follow the analysis in \cite{AlvarezGaume:2011rk,Alvarez-Gaume:2013iqa} and to explore solutions we follow the procedures given in \cite{Antonelli:2019dqv, Hirschmann_1997}. Let us describe very briefly the properties of the self-similar solutions. Basically, one  obtains a system of ordinary differential equations (ODEs) 
\begin{align}
    b'(z) & = B(b(z),f(z),f'(z)),\quad 
    f''(z)  = F(b(z),f(z),f'(z))\,. \label{eq:unperturbedfpp}
\end{align}
These equations have five singularities: 
\begin{eqaed}
    z & =  0\\
    z & =  z_+>0\,,\quad b(z_+) = z_+\\
    z & =  \pm \infty \\
    z & =  z_-<0\,,\quad b(z_-) = - z_-
\end{eqaed}
The point $z=\pm 0$ is related to the axis $r=0$ and the regularity condition can be readily applied. Assuming that the scalars are regular across this axis and using time re-scaling, one can obtain
\begin{equation}
    f'(0) =0,\quad  b(0) = 1\,.
\end{equation}
which provide three real boundary conditions\footnote{The point $z=\infty$ is related to the surface $t=0$. We used the change of variable for the fields $f(z)$, $b(z)$ and showed that nothing special happens on this surface (see the appendices in \cite{Alvarez-Gaume:2013iqa}).}.
Using the residual symmetry for the e.o.m.'s mentioned in the last section, one would be able to eliminate one degree of freedom from the complex number $f(0)$. Ultimately we let
\begin{equation}\label{eq:zeroffixing}
        f(0) = \left\{\begin{array}{l l l}
        x_0 & \text{elliptic}       & 0<x_0<1 \\
        i x_0 & \text{parabolic} & 0<x_0\\
        1+i x_0 & \text{hyperbolic} & 0<x_0
    \end{array}\right.
\end{equation}
so that the problem is reduced to the determination of just two real parameters, $\omega$ and $x_0$.

The singularities $b(z_{\pm})=\pm z_{\pm}$ are related to backward and forward light cones of the spacetime origin and also correspond to the surfaces where the homothetic Killing vector becomes null. The solution is smooth across $b(z_+)=z_+$, while the forward cone $b(z_-)=-z_-$ is the Cauchy horizon of the spacetime and in this region we need to have the continuity of $f$, $b$.
%\framebox{what do you mean here?}.
Now one needs to consider the relevant part for $z$ that embeds the infinite past and exists between the following two singularities
\begin{equation}
    z = 0,
    \end{equation}
    \begin{equation}
    z = z_+\,,\quad b(z_+) = z_+\,.
\end{equation}
The surface $z=z_+$ is also a coordinate singularity, and the  field $\tau(t,r)$ should be regular across it, which actually means that $f''(z)$ must remain finite as $z\rightarrow z_+$. We also note that the vanishing of this divergent part of $f''(z)$ is indeed a complex-valued constraint at $z_+$
\begin{equation}\label{eq:unpconstraint}
    \mathbb{C} \,\ni\, G(b(z_+), f(z_+), f'(z_+))=0\,.
\end{equation}
The explicit form of the constraint $G(b(z_+), f(z_+), f'(z_+))$ for all three different cases can be found in Section 4.1.2 of \cite{Antonelli:2019dqv}. 

Therefore, our numerical procedure is as follows. We first determine $f(0)$ from $x_0$ according to  \eqref{eq:zeroffixing}. We then use the boundary conditions at $z=0$. We also start integrating forward the e.o.m.'s from a small $z_0$ to avoid the singular point $z=0$. We stop the integrations once $b(z)-z$ reaches a lower level $\delta$ that is also positive and small. The value of $z$ where the crossing occurs is $z_+$. Finally we make use of numerical solution to explore  $f(z_+)$, $f'(z_+)$, $b(z_+)$ as well as the result of the constraint $G$ introduced in \eqref{eq:unpconstraint}. Therefore, we are left out with two distinct constraints (which are the real and imaginary parts of $G$)  with just two unknown parameters $(\omega,x_0)$. Thus, we draw the curves where $\Re G$ and $\Im G$ vanish in the plane of $(\omega,x_0)$ and essentially look for their intersections. Hence, using a starting point, the root-finding reveals the locations of the roots.
Therefore, ODE's can be entirely solved which have a discrete solution set. For further numerical explanations, see Section 4 of \cite{Antonelli:2019dqv}. In order to provide some background, we now briefly revisit the CSS solutions in lower dimensions, %\cite{Antonelli:2019dqv}
before generalizing them to six and seven dimensions for all three conjugacy classes.
\subsection{Results}
The self-similar solutions in four and five dimensions were recently determined in \cite{Antonelli:2019dqv} by two unknown parameters of $(\omega,x_0)$, and we also explored the precise location of the $z_+$ singularity.
One can construct the solutions integrating numerically the CSS e.o.m.'s. For the sake of completeness, here we briefly mention those solutions and then start exploring higher--dimensional solution sets in parameter space with their figures in six and seven dimensions for all distinct cases of the elliptic, parabolic and hyperbolic classes.
\subsubsection{Self-similar solutions for $d=4,5$ elliptic class}
In \cite{Antonelli:2019dqv} the curves of vanishing real and imaginary parts of the constraint $G(\omega,|f(0)|)$ for the $d=4$ elliptic case were drawn. Only one solution at the intersection was obtained, whose parameters are
\begin{eqaed}
    w & = 1.176, \quad \abs{f(0)}= 0.892, \quad z_+= 2.605
\end{eqaed}
Note that this single solution was also explored in \cite{Eardley:1995ns,AlvarezGaume:2011rk}.
%and there is no any other solution outside of the range of our plot.
\begin{figure}[H]
   \centering
 \includegraphics[width=2.5in]{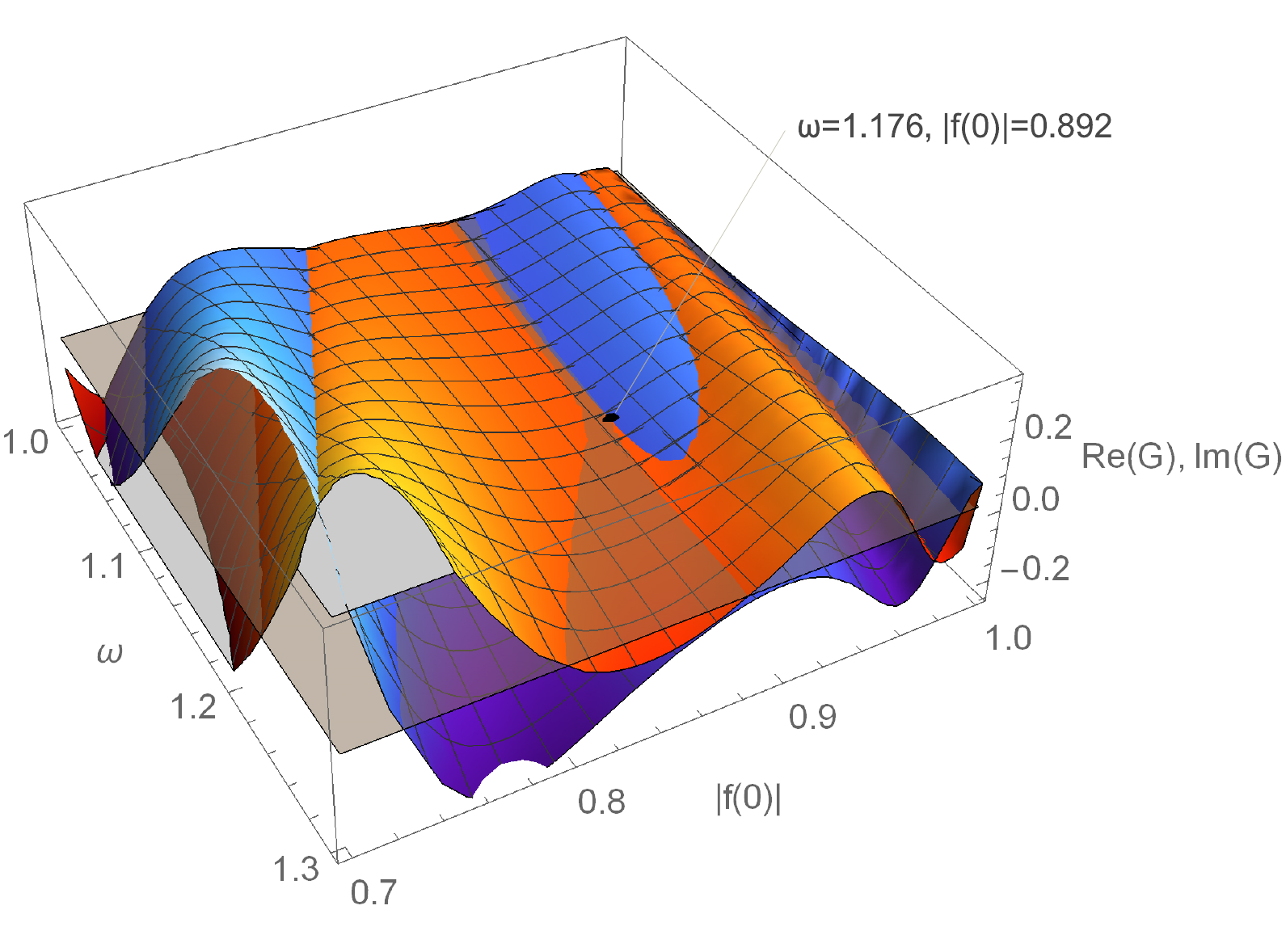}
   \caption{Three dimensional plot of the solution for elliptic class in four dimensions. The real part of the constraint $G(\omega,|f(0)|)$ is plotted in orange, the imaginary part in blue. The plane of vanishing $\Re (G),\Im (G)$ is shown in gray, and this is where the intersection should be looked for.}
    \label{4Del3d}
\end{figure}
On the other hand, for the five dimensional elliptic case three distinct solutions ($\alpha$, $\beta$ and $\gamma$) were found \cite{ Antonelli:2019dqv}. We arrange them in order of increasing $\omega$ and represent those three solutions in Table \ref{tab:5Delliptic}.
%Notice that we did not find any other solutions outside that specific range.
\begin{table}[ht]
\begin{center}
\bgroup
\def\arraystretch{2}
\begin{tabular}{|c|c|c|c|}\hline
    Solution & $w$ & $\abs{f(0)}$ & $z_+$ \\
    \hline
    $\alpha$ & $0.999$ & $0.673$ & $1.246$\\
    $\beta$ & $1.680$ & $0.644$ & $1.397$ \\
    $\gamma$ & $2.304$ & $0.700$ & $1.694$\\
    \hline
\end{tabular}
\egroup
\caption{Solutions for the elliptic class in five dimensions.}
\label{tab:5Delliptic}
\end{center}
\end{table}
\section{Solutions for $d=6$ elliptic class}
\begin{figure}[h!]
\centering
\begin{subfigure}[t]{.5\textwidth}
 \centering
 \includegraphics[width=2.3in]{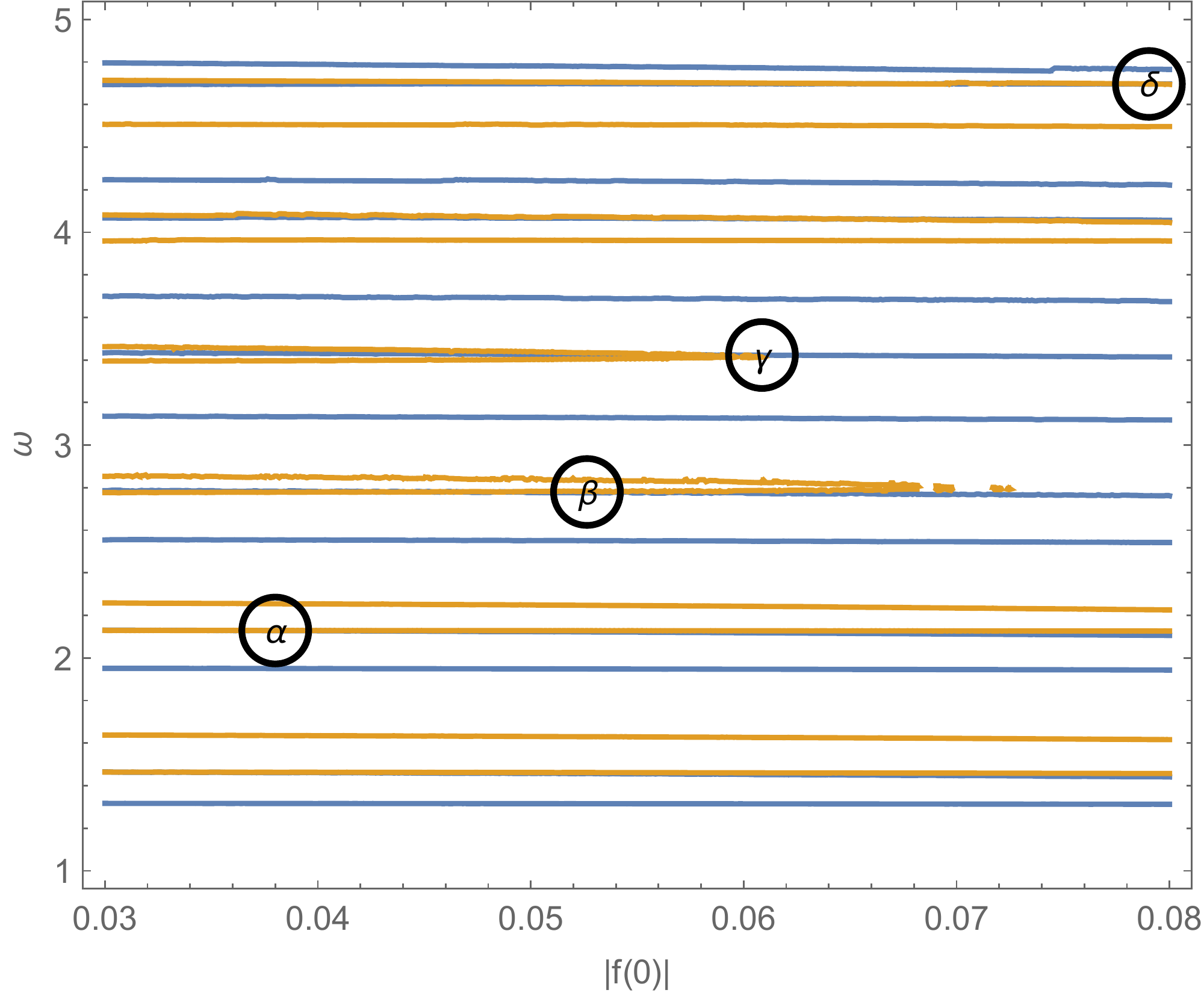}
 \caption{}
 \label{6dallsolselliptic}
\end{subfigure}%
\begin{subfigure}[t]{.5\textwidth}
  \centering
  \includegraphics[width=2.35in]{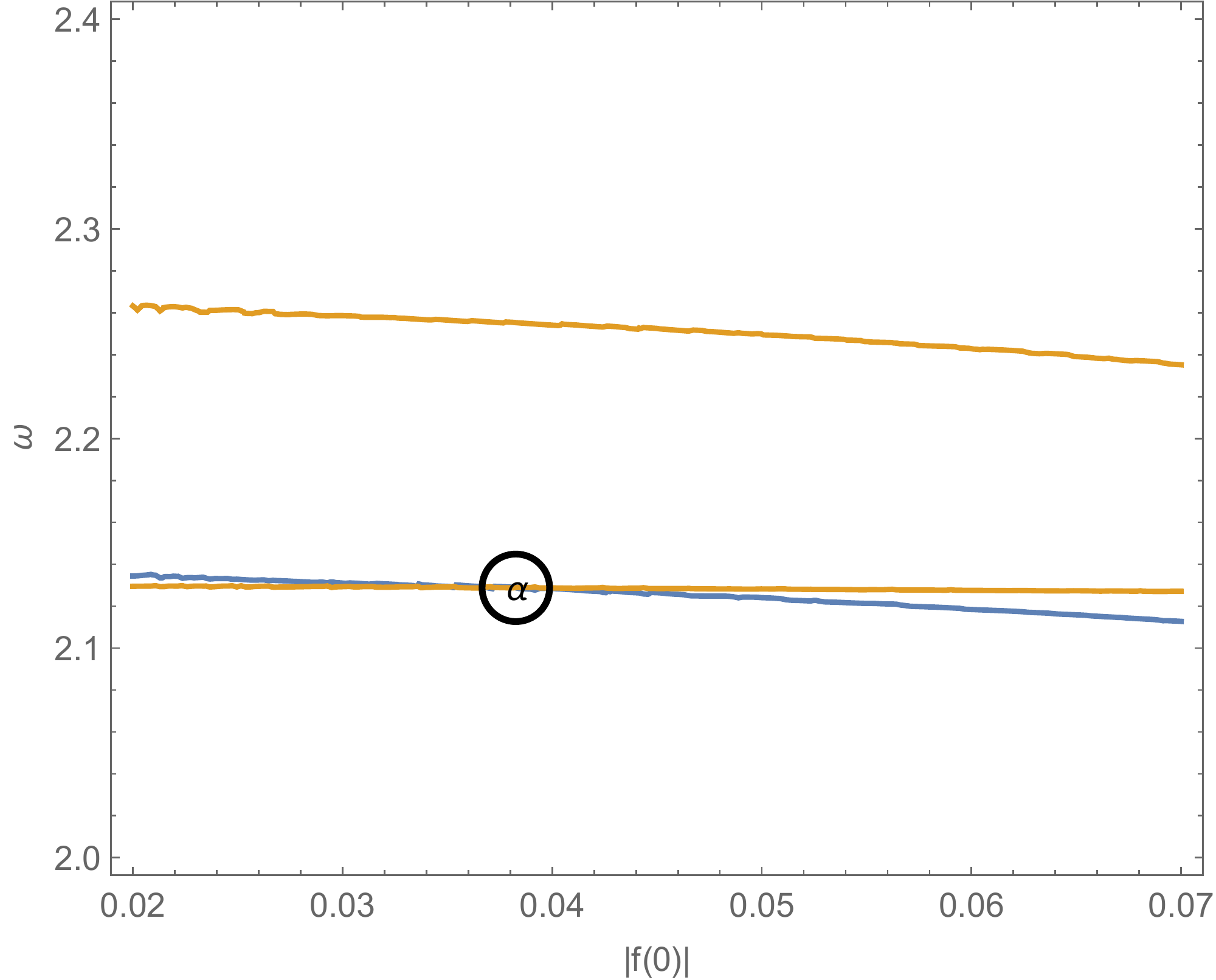}
  \caption{}
  \label{6d1stsolelliptic}
\end{subfigure}
\caption{(a) Curves of vanishing real and imaginary parts of $G(\omega,\abs{f(0)})$ in the $d=6$ elliptic case. Four solutions were found in order of increasing $\omega$ and $|f(0)|$. (b) The curves of vanishing real and imaginary parts of $G(\omega,\abs{f(0)})$ in the $d=6$ elliptic case for $\alpha$ branch.}
\label{6dsolelliptic}
\end{figure}

In this section we would like to explore higher dimensional self-similar solutions for the elliptic class. 
We determine the curves of vanishing real and imaginary parts of $G(\omega,|f(0)|)$ for a wide range of $(\omega,|f(0)|)$ for the $d=6$ elliptic case. We were able to identify four intersections corresponding to four solutions that are being called $\alpha$, $\beta$, $\gamma$ and $\delta$, in order of increasing $\omega$ and $|f(0)|$, as depicted in \ref{6dallsolselliptic}. We have taken  $\beta$ and $\gamma$ solutions with $(G \sim 10^{-11}-10^{-12}$) and with confidence. Note that, due to the small value of $|f(0)|$, the quality of the $\alpha$ and $\delta$ solutions is of order $(G \sim 10^{-4}-10^{-5}$). We illustrate these four solutions accordingly. No other solution  was found outside this range.
Using Figure~\ref{6d1stsolelliptic}, we are able to identify the first branch solution in the $d=6$ elliptic class, which we called $\alpha$ solution and whose parameters are\footnote{ We keep track of solutions in higher dimensions up to four decimal places.}:
\begin{eqaed}
    w & =2.1287, \quad  \abs{f(0)}= 0.0383, \quad z_+ = 1.0007 \ ,
\end{eqaed}
\begin{table}[ht]
\begin{center}
\bgroup
\def\arraystretch{2}
\begin{tabular}{|c|c|c|c|}\hline
    Solution & $w$ & $\abs{f(0)}$ & $z_+$ \\
    \hline
    $\alpha$ & $2.1287$ & $0.0383$ & $1.0007$\\
    $\beta$ & $2.7797$ & $0.0526$ & $1.0021$ \\
    $\gamma$ & $3.4232$ & $0.0609$ & $1.0033$\\
    $\delta$ & $4.6968$ & $0.0790$ & $1.0070$\\
    \hline
\end{tabular}
\egroup
\caption{Solutions for the elliptic class in six dimensions.}
\label{tab:6Delliptic}
\end{center}
\end{table}
and we investigated all the other branches in a similar fashion. The solutions  for the six-dimensional elliptic case are summarized in Table \ref{tab:6Delliptic}.
\section{Solutions for $d=7$ elliptic class}
For the seven-dimensional elliptic class we have been able to identify three solutions at the intersections that are called $\alpha$, $\beta$, $\gamma$, in order of increasing $\omega$. We have identified  all these three solutions with $(G \sim 10^{-9}-10^{-11}$) with a good confidence. We represent those solutions in Table \ref{tab:7Delliptic}. 
\begin{table}[ht]
\begin{center}
\bgroup
\def\arraystretch{2}
\begin{tabular}{|c|c|c|c|}\hline
    Solution & $w$ & $\abs{f(0)}$ & $z_+$ \\
    \hline
    $\alpha$ & $0.0023$ & $0.9996$ & $1.8032$\\
    $\beta$ & $0.0078$ & $0.9990$ & $2.0440$ \\
    $\gamma$ & $0.0153$ & $0.9989$ & $2.8275$\\
    \hline
\end{tabular}
\egroup
\caption{Solutions for the elliptic class in seven dimensions.}
\label{tab:7Delliptic}
\end{center}
\end{table}

\begin{figure}[H]
    \centering
    \includegraphics[width=2.3in]{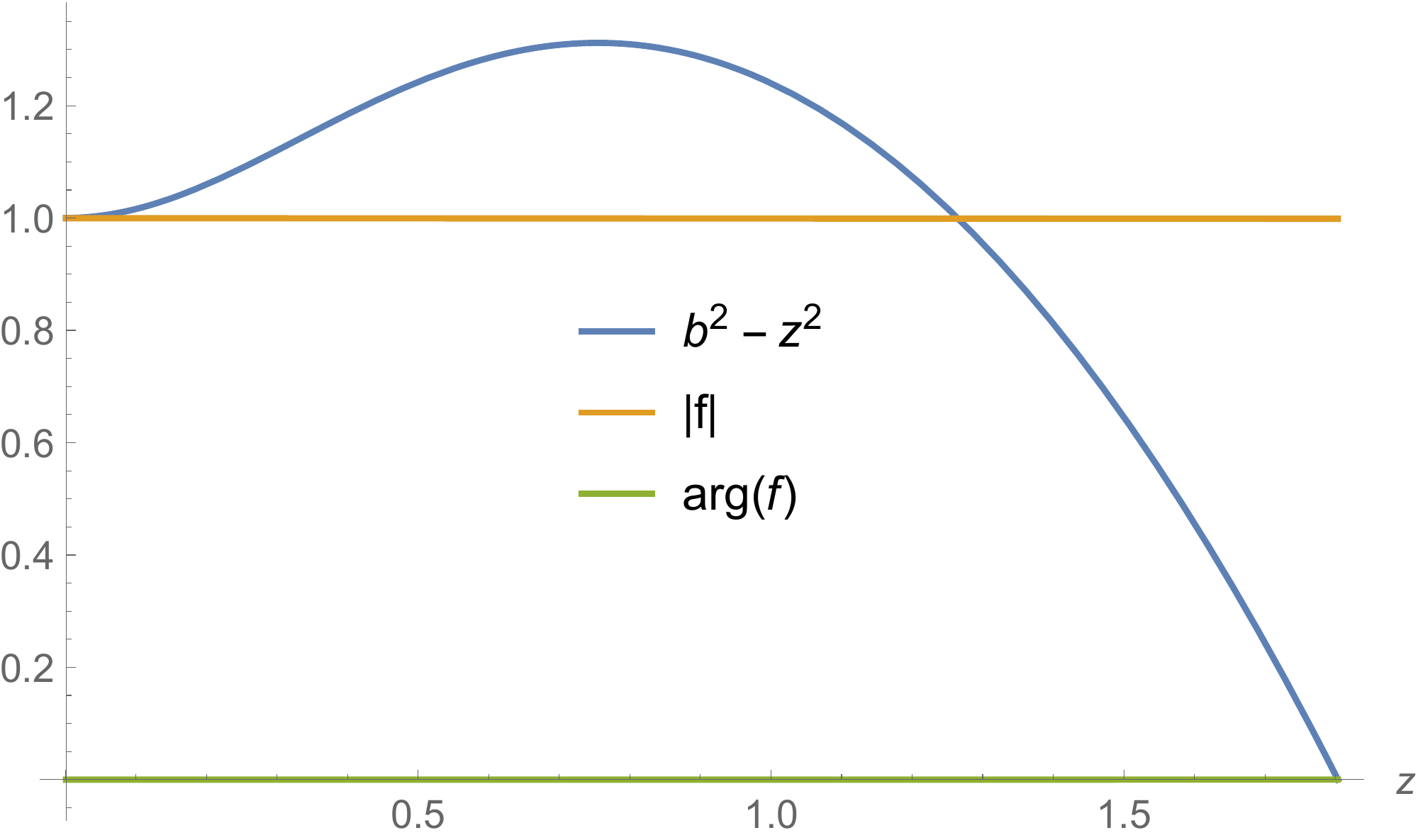}
    \caption{Profile of $\alpha$ solution for elliptic case in seven dimensions.}
    \label{7dalpha}
\end{figure}

For the sake of brevity we just display the profile of $\alpha$ solution in Figure~\ref{7dalpha}.
%Notice that no solution was found for $1 \leq \omega \leq 9$. %as depicted in below.
\section{Solutions for $d=4,5$ parabolic class}
Due to extra symmetry for the parabolic class (see \eqref{parabolic_rescaling}), if the parameters $(\omega,\Im f(0))$ correspond to a solution, so do $(K \omega,K\Im f(0))$, since both e.o.m.'s and the $G(\omega,\Im f(0))$ are invariant under this scaling. Hence, the ratio $\omega/\Im f(0)$, is the only real unknown parameter for this parabolic class, and one must search for zeroes of $G(\omega,\Im f(0))$ over just one real parameter $\omega/\Im f(0)$. Here we draw this complex function over $\omega$ for the particular $\Im f(0)=1$. Notice that, if a root $\omega^*$ existed, it would produce a continuous ray of solutions $(\omega,\Im f(0)) = (K \omega^*, K)$.

\begin{figure}[H]
    \centering
    \includegraphics[width=2.5in]{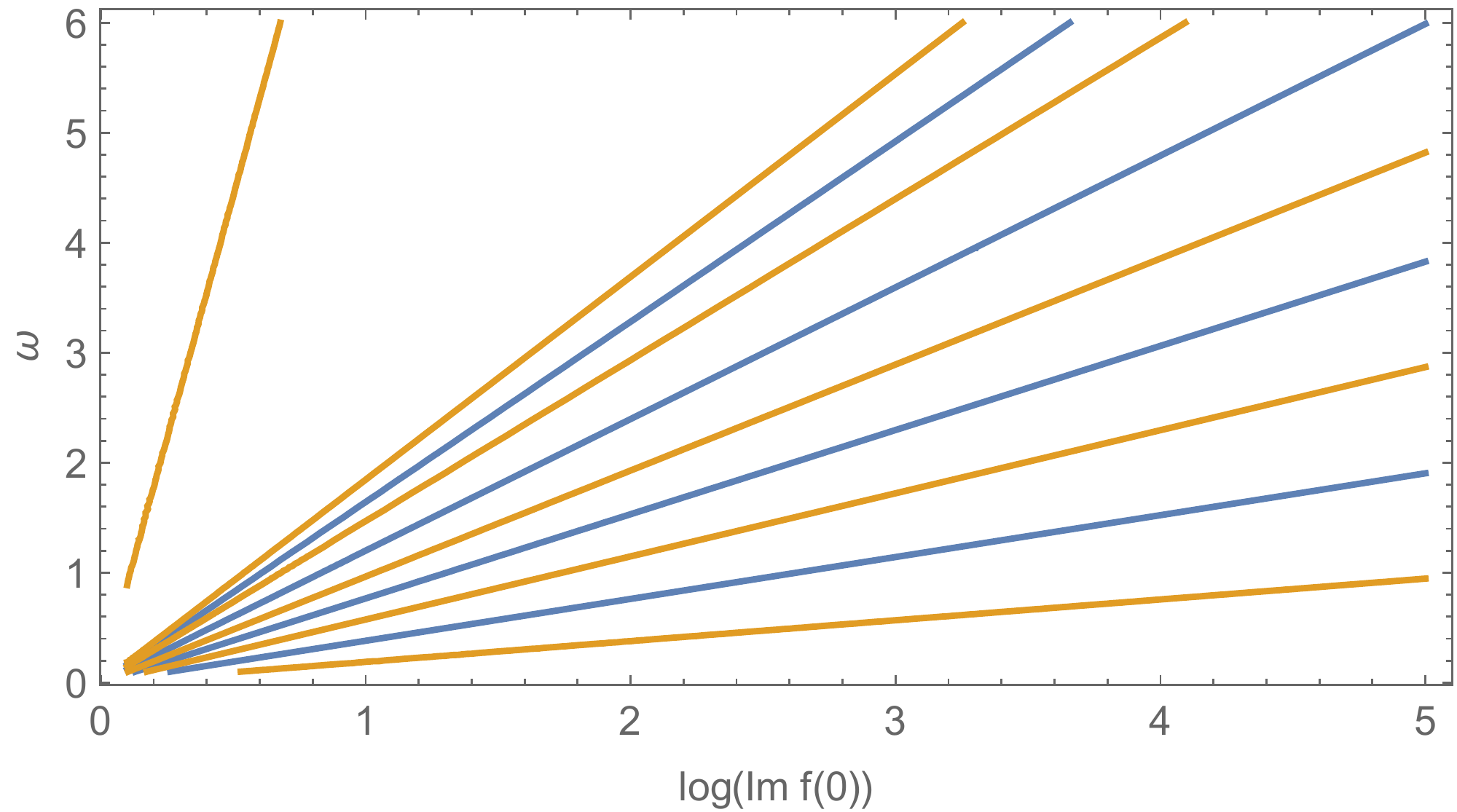}
    \caption{A two-dimensional plot of the zeroes of the real and imaginary parts of $G(\omega,\Im f(0))$ in five dimensions.}
    \label{51dparabolic}
\end{figure}

The plots of absolute value and both real and imaginary parts of $G(\omega,1)$ over $\omega$ were displayed in \cite{Antonelli:2019dqv}, and no zeroes were notified for $\omega > 0$ in the four--dimensional parabolic class. In Figure \ref{51dparabolic} we show a two-dimensional plot of the zeroes of the real and imaginary parts of $G(\omega,\Im f(0))$ in five dimensions, which clarifies the degeneracy related to the extra scaling symmetry of the parabolic class.
\begin{figure}[H]
    \centering
    \includegraphics[width=2.5in]{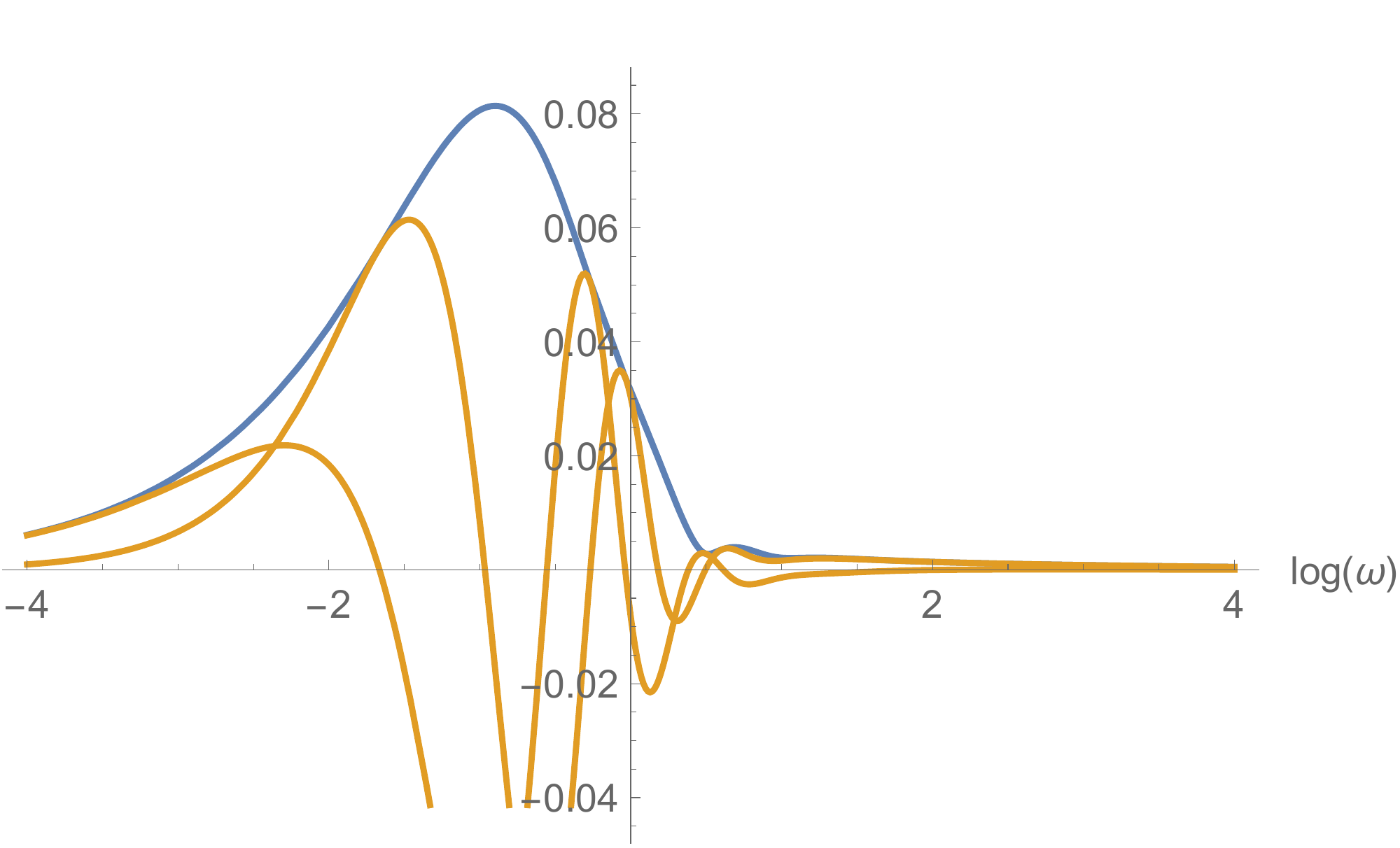}
    \caption{The plot of absolute value (blue), real and imaginary parts (orange) of $G(\omega,1)$ in the $d=5$ parabolic case.}
    \label{5dpara2}
\end{figure}
In Figure \ref{5dpara2} for the five--dimensional parabolic case one might notice a very small value of $\abs{G}$ around a particular $\omega$ as the following solution ray:
\begin{eqaed}
   \abs{G} \sim 0.006,\quad   \omega \sim 1.65\ .
\end{eqaed}
We are not able to exclude these small values as a solution ray in five dimensions that may not have been identified due to numerical errors. Indeed, this might be a possible solution for an over-determined configuration.
\section{Solutions for $d=6,7$ parabolic class}
In order to see whether or not there are possibly additional solutions for the parabolic class in higher dimensions, here we illustrate in two-dimensional plots the zeroes of the real and imaginary parts of $G(\omega,\Im f(0))$, which displays the degeneracy caused by a scaling invariance of the parabolic class. It was not possible to identify solution rays, in both six and seven dimensions, and we are tempted to conjecture that there are also no solutions for the parabolic case in higher dimensions. 

\begin{figure}[H]
    \centering
    \includegraphics[width=2.5in]{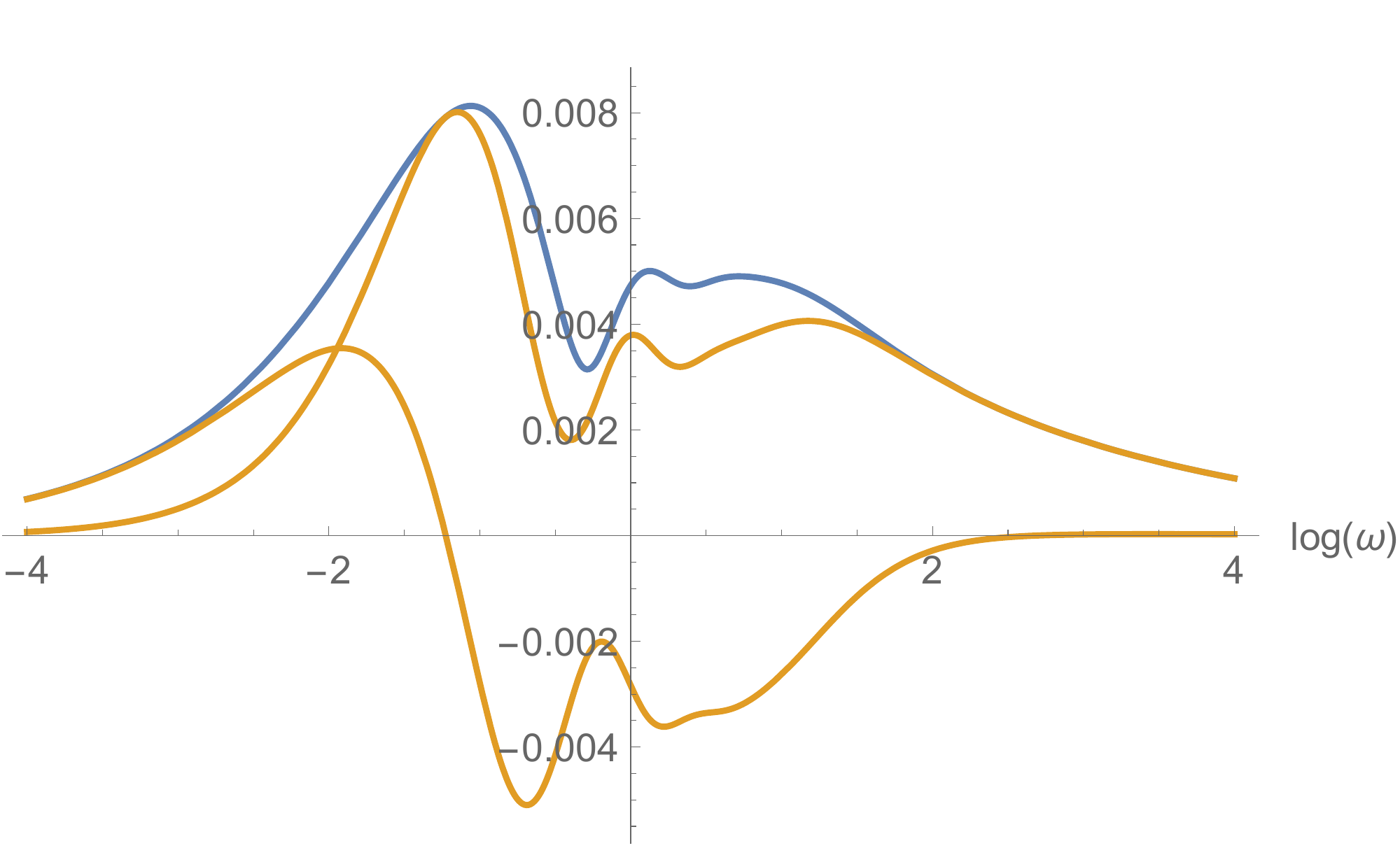}
    \caption{The plots of absolute value (blue) and real and imaginary parts (orange) of $G(\omega,1)$ over $\omega$ in the $d=6$ parabolic case. We cannot identify any  zero for $\omega > 0$.}
    \label{fig:4dparabolic}
\end{figure}
\begin{figure}[H]
    \centering
    \includegraphics[width=2.5in]{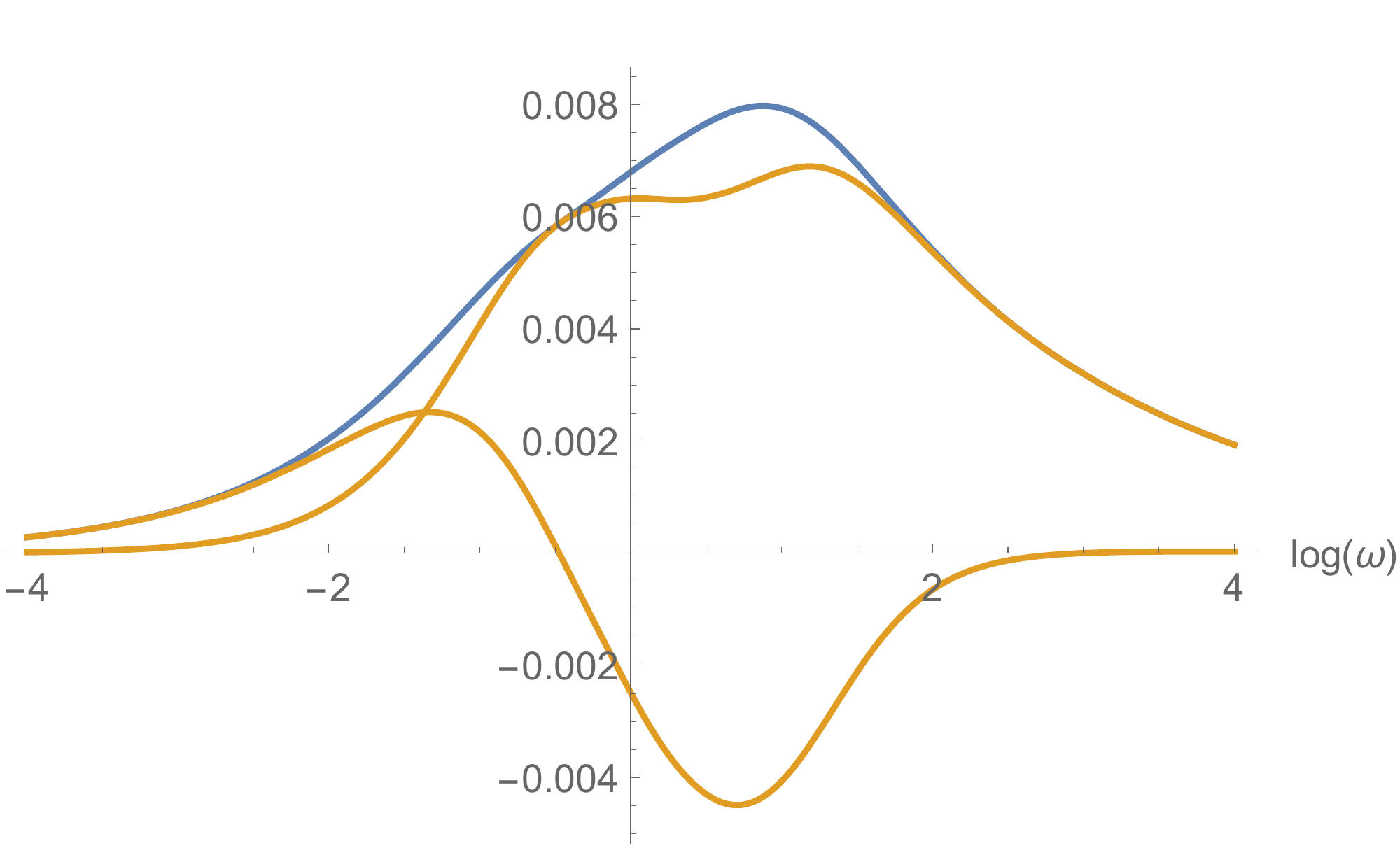}
    \caption{The plots of absolute value (blue) and real and imaginary parts (orange) of $G(\omega,1)$ over $\omega$ in the $d=7$ parabolic case. Again, we cannot identify any zero for $\omega > 0$.}
    \label{fig:4dparabolic}
\end{figure}
\section{Solutions for $d=4,5$ hyperbolic class}
\begin{figure}[H]
    \centering
    \includegraphics[width=3in]{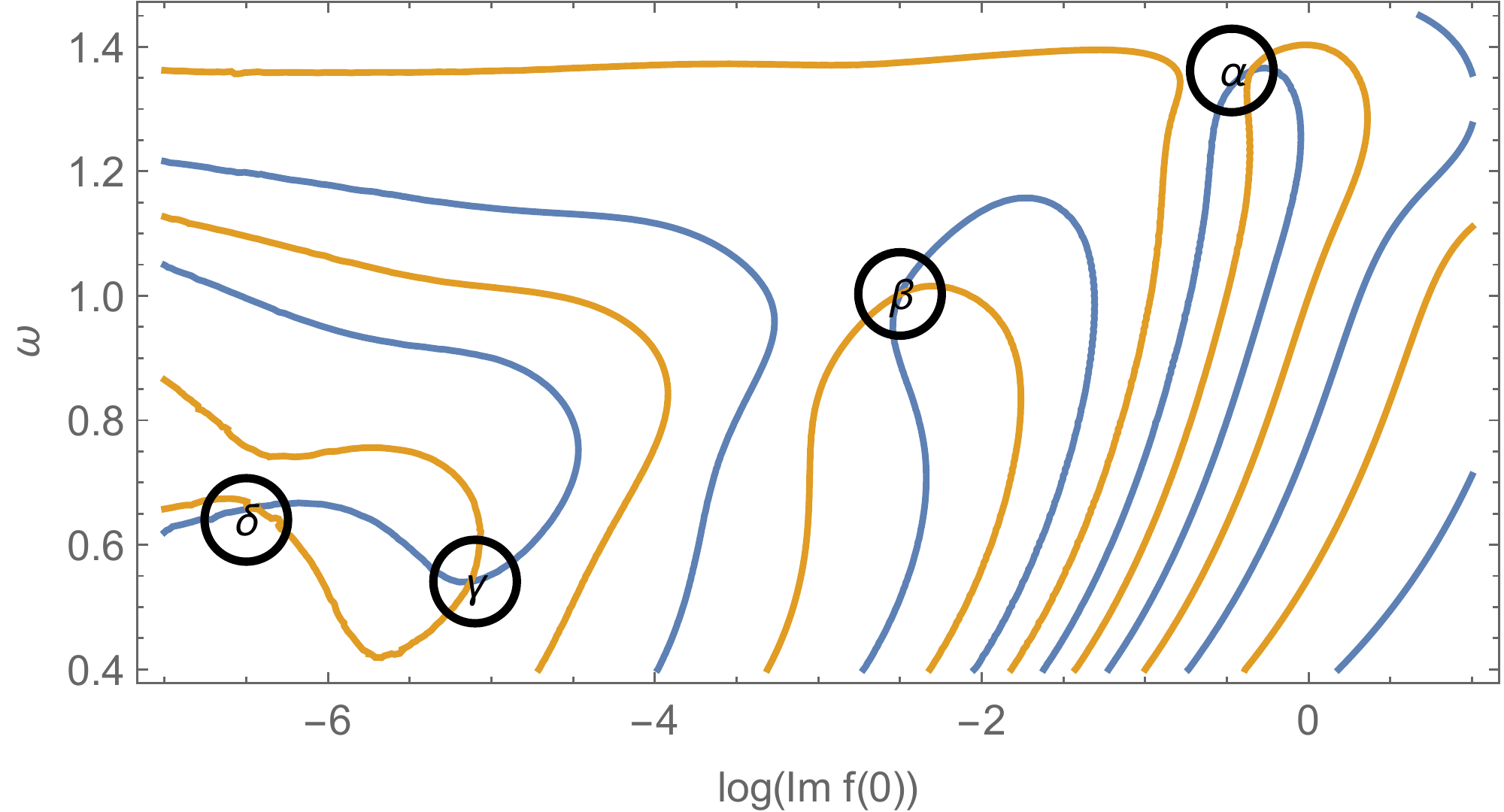}
    \caption{The curves of vanishing real and imaginary parts of $G(\omega,\Im f(0))$ for $d=4$ hyperbolic case.}
    \label{fig:4hyperbolic}
\end{figure}

\begin{figure}[H]
    \centering
    \includegraphics[width=2.5in]{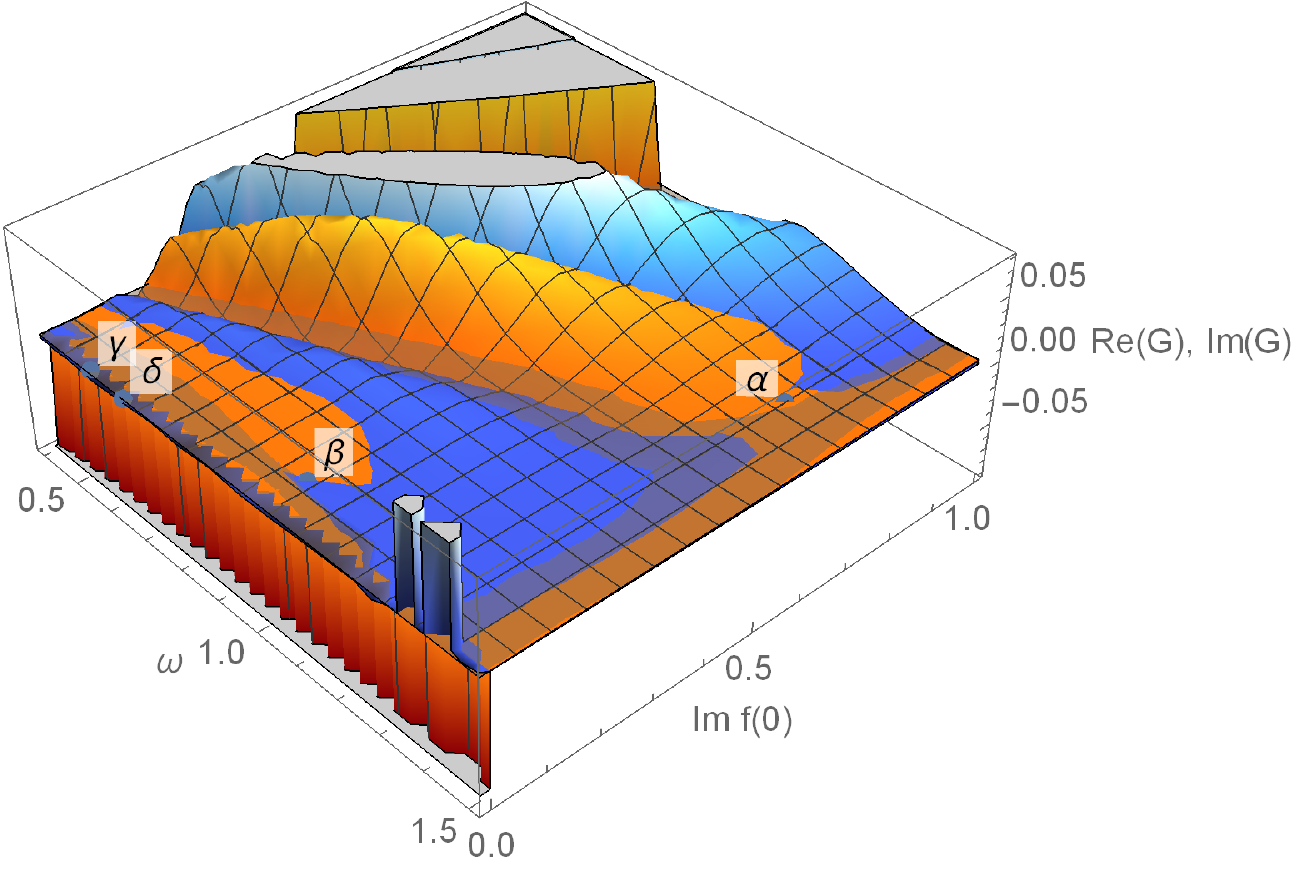}
    \caption{Three--dimensional plot of the solution for the hyperbolic class in four dimensions. The real part of the constraint $G(\omega,\Im f(0))$ is plotted in orange, the imaginary part in blue. The vanishing plane $\Re (G),\Im (G)$ is shown in gray. The solutions lie at the intersection of these three surfaces.}
    \label{4Dhy3d}
\end{figure}

\begin{figure}[H]
   \centering
   \includegraphics[width=2.5in]{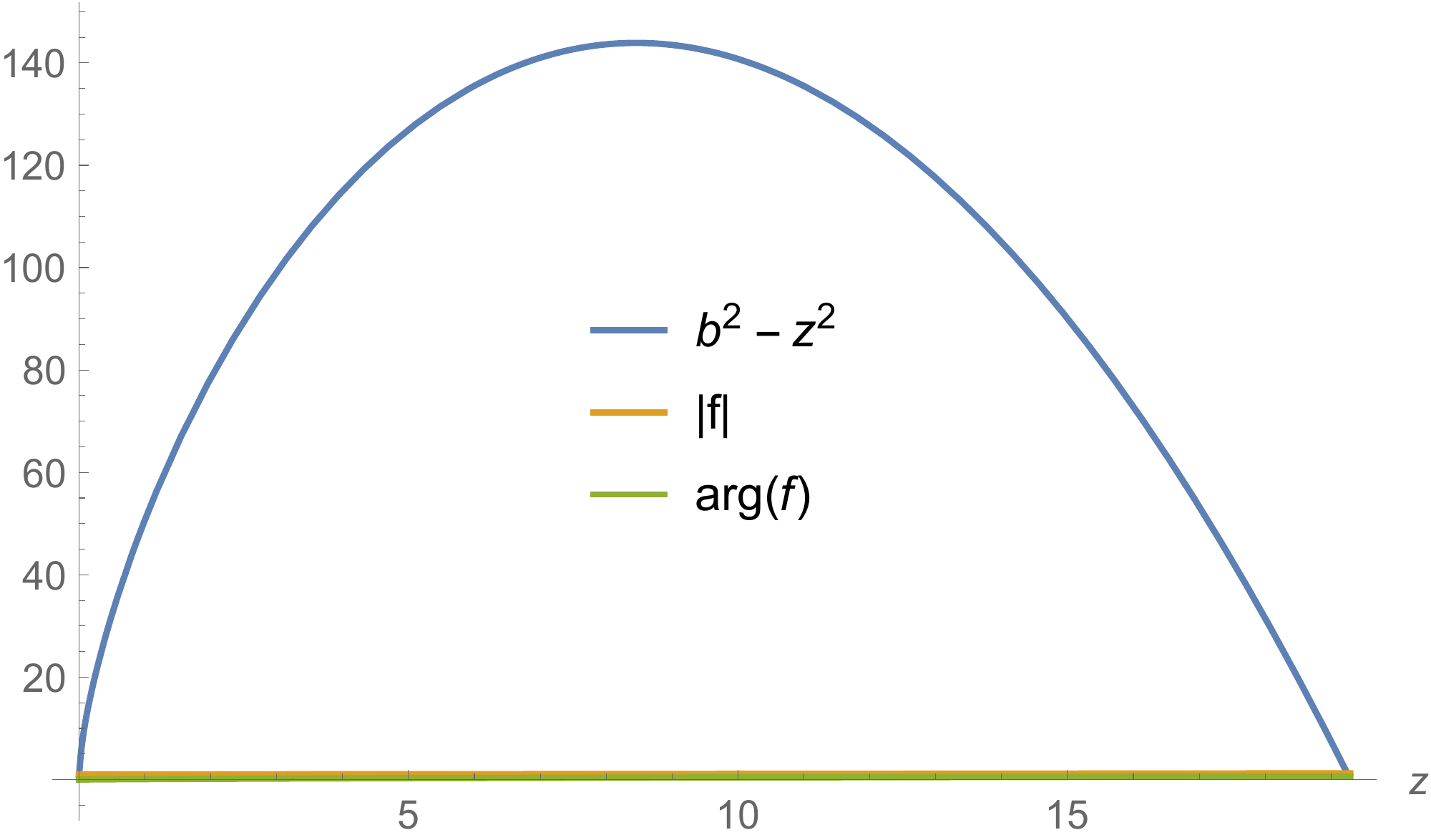}
    \caption{The profile of $\delta$ branch for hyperbolic class in four dimensions.}
    \label{fig:4thsol4dhyper}
\end{figure}

Four distinct solutions for four dimensions in the hyperbolic case were explored in  \cite{Antonelli:2019dqv}, in order of decreasing $\Im f(0)$, as depicted in Figure \ref{fig:4hyperbolic}.
As was argued, using a root-finding procedure one can also investigate the fourth intersection, which is called $\delta$ whose parameters are given by:
\begin{eqaed}
    w & =0.6404, \quad  \Im f(0) = 0.0015, \quad  z_+ &= 19.2923
\end{eqaed}
The three dimensional plot of the constraint $G(\omega,\Im f(0))$ is also shown in Figure \ref{4Dhy3d}.
The profile of $\delta$ solution is shown in Figure \ref{fig:4thsol4dhyper}.
The solutions in four and five dimensions are summarised in Table \ref{tab:4Dhyper} and in Table \ref{tab:5Dhyper}.
\begin{figure}[H]
\centering
\begin{subfigure}[t]{.5\textwidth}
  \centering
  \includegraphics[width=.99\linewidth]{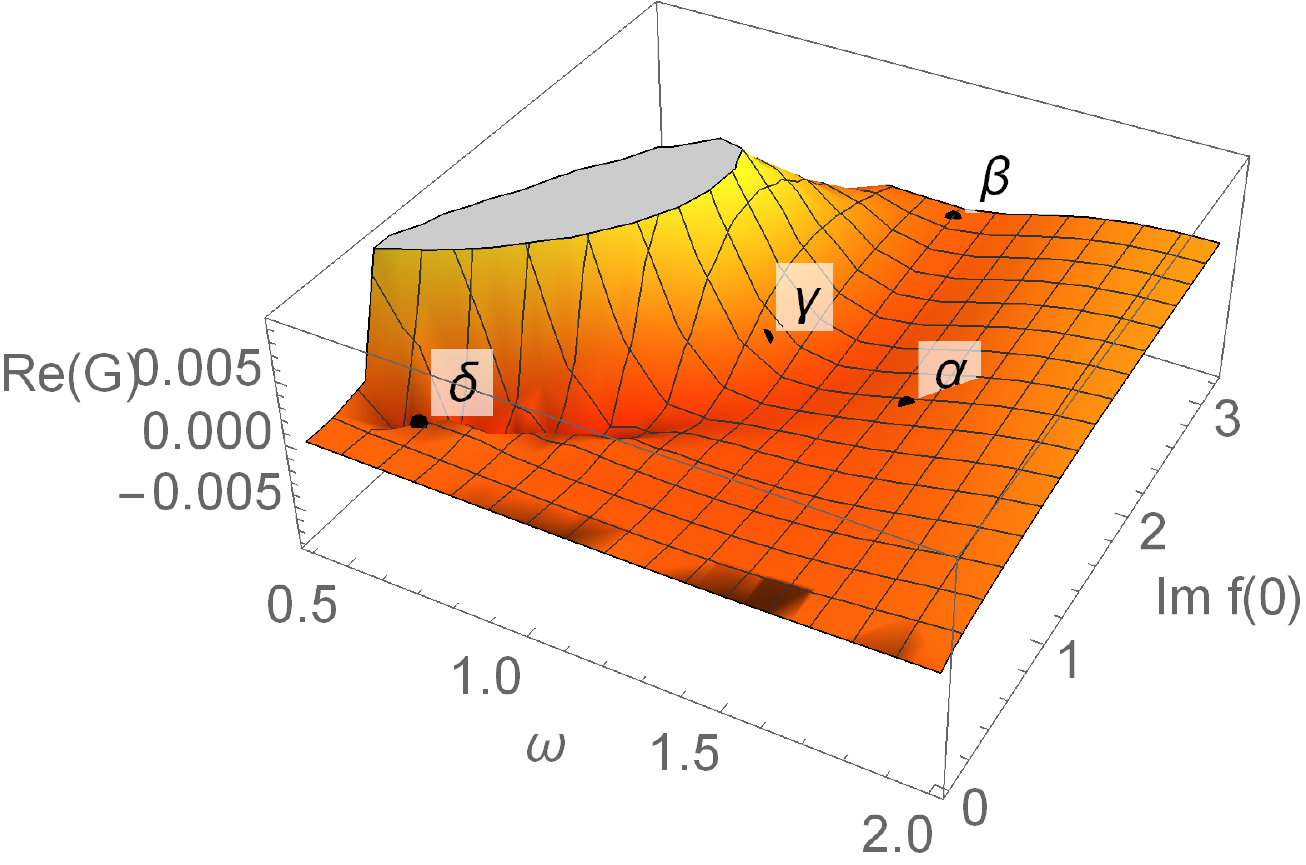}
  \caption{Real part of $G(\omega,\Im f(0))$.}
  \label{}
\end{subfigure}%
\begin{subfigure}[t]{.5\textwidth}
  \centering
  \includegraphics[width=.99\linewidth]{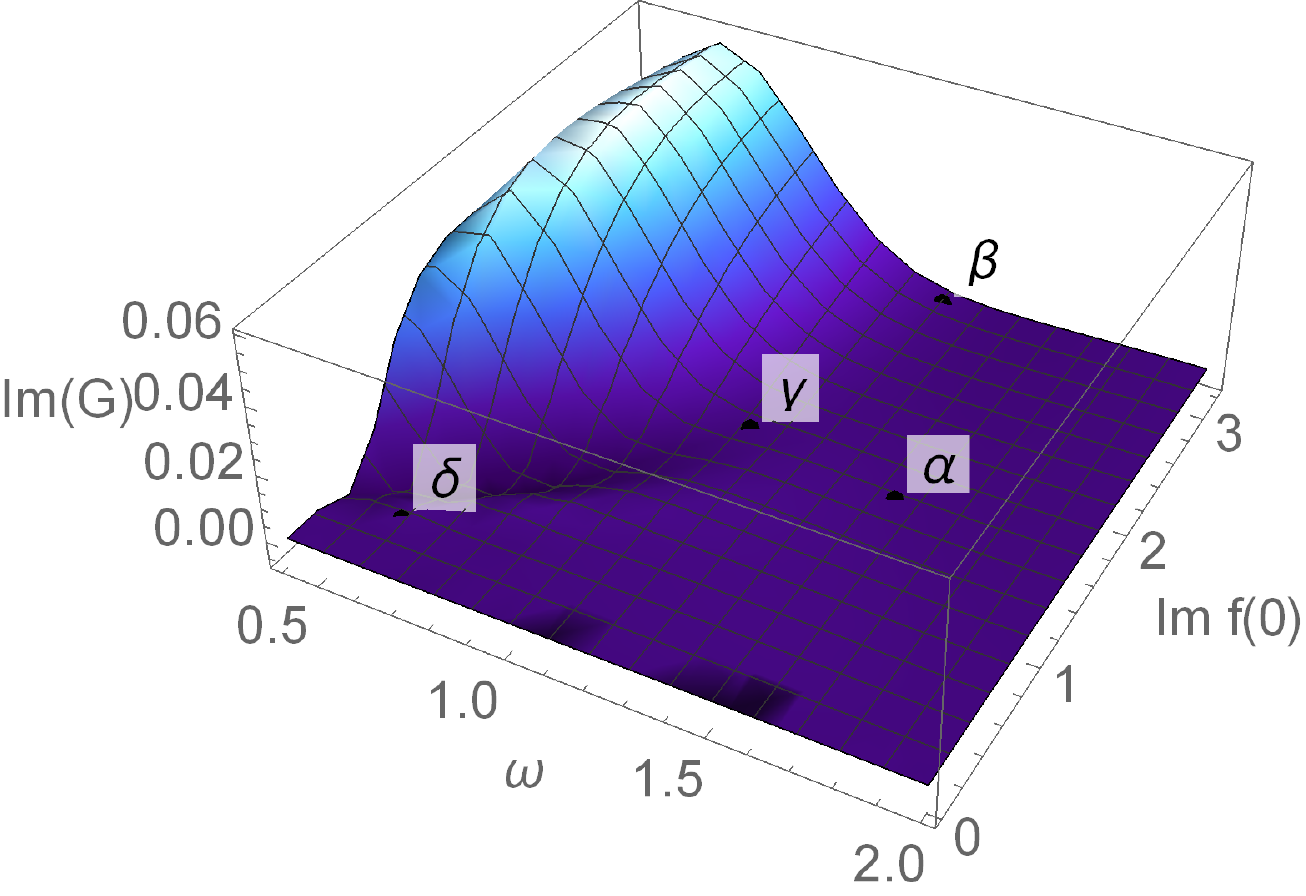}
  \caption{Imaginary part of $G(\omega,\Im f(0))$.}
  \label{}
\end{subfigure}
\caption{Three dimensional plots of the constraint function $G(\omega,\Im f(0))$ for the hyperbolic solutions in five dimensions.}
\label{5ddDhyper}
\end{figure}

%SIDE BY SIDE
\begin{table}[h!]
\raggedright
\begin{minipage}{0.42\linewidth}
\begin{center}
\bgroup
\def\arraystretch{2}
 \begin{tabular}{|c|c|c|c|}\hline
    Solution & $w$ & $\Im f(0)$ & $z_+$ \\ \hline
     $\alpha$ & $1.362$ & $0.708$ & $1.440$\\
     $\beta$ & $1.003$ & $0.0822$ & $3.29$ \\
     %\rowcolor{light-gray}
     $\gamma$ & $0.541$ & $0.0059$ & $8.44$\\
     %\rowcolor{light-gray}
     $\delta$ & $0.6404$ & $0.0015$ & $19.2923$\\ 
     \hline
 \end{tabular}
 \egroup
 \captionof{table}{4d hyperbolic solutions.}
 \label{tab:4Dhyper}
 \end{center}
\end{minipage}
\hspace{0.1\linewidth}
\begin{minipage}{0.42\linewidth}
\begin{center}
\bgroup
\def\arraystretch{2}
\begin{tabular}{|c|c|c|c|}\hline
    Solution & $w$ & $\Im f(0)$ & $z_+$ \\ \hline
    $\alpha$ & $1.546$ & $1.555$ & $1.254$\\
    $\beta$ & $1.305$ & $3.086$ & $1.129$\\
    $\gamma$ & $1.125$ & $1.705$ & $1.109$\\
    $\delta$ & $0.588$ & $0.364$ & $1.156$\\
    \hline
 \end{tabular}
 \captionof{table}{5d hyperbolic solutions.}
 \label{tab:5Dhyper}
 \egroup
\label{tab:4D-5Dhyper}
\end{center}
\end{minipage}
\end{table}

\section{Solutions for $d=6,7$ hyperbolic class}

For the hyperbolic class, especially in higher dimensions, we were able to obtain two solutions, in both six and seven dimensions.  They are displayed in Tables \ref{tab:6Dhyper} and \ref{tab:7Dhyper}, in order of decreasing $\omega$ and $\Im f(0)$. The output of the root-finding procedure is accurate, and hence we have checked that the $\alpha$ and $\beta$ solutions can be identified with $(G \sim 10^{-11}-10^{-12})$ with good precision. The source of inaccuracy in the solutions can be ascribed to the magnitude of the regularisation parameters $z_0$ and $\delta$\footnote{So far we have set $z_0=0.001$ and $\delta=10^{-4}$. Instead, for the hyperbolic class in six and seven dimensions we have chosen $z_0=0.0001$ and $\delta=10^{-5}$. We have also checked that our previous results in lower dimensions are not affected if we change these parameters.}.

\vskip.2in

\begin{table}[h!]
\raggedright
\begin{minipage}{0.42\linewidth}
\begin{center}
\bgroup
\def\arraystretch{2}
\begin{tabular}{|c|c|c|c|}\hline
    Solution & $w$ & $\Im f(0)$ & $z_+$  \\ 
    \hline
    $\alpha$ & $0.0100$ & $0.0070$ & $1.0901$ \\
    $\beta$ & $0.0063$ & $0.0038$ & $1.1201$ \\
    \hline
\end{tabular}
\egroup
\captionof{table}{6d hyperbolic solutions.}
\label{tab:6Dhyper}
\end{center}
\end{minipage}
\hspace{0.1\linewidth}
\begin{minipage}{0.42\linewidth}
\begin{center}
\bgroup
\def\arraystretch{2}
\begin{tabular}{|c|c|c|c|c|}\hline
    Solution & $w$ & $\Im f(0)$ & $z_+$   \\ \hline
    $\alpha$ & $0.0021$ & $0.0014$ & $1.0828$  \\
    $\beta$ & $0.0017$ & $0.0011$ & $1.0952$  \\
    \hline
\end{tabular}
\captionof{table}{7d hyperbolic solutions.}
\label{tab:7Dhyper}
\egroup
\label{tab:6D-7Dhyper}
\end{center}
\end{minipage}
\end{table}

Notice that, due to the small values of $\Im f(0)$, we can not exclude the presence of further solutions for the hyperbolic class in higher dimensions. The numerical accuracy decreases as the dimension increases and we trust the solution parameters only up to four decimal places. Indeed, for the higher dimensional hyperbolic class we could not rely on the graphical representations to get hints as to where the intersections of $\Re (G), \Im (G)$ may lie. Still, the numerical procedure does discover some roots with good precision and we are confident about them.

\section{Conclusions}
In this paper we have shown that there are various spherically symmetric self-similar collapse solutions for the Einstein-axion-dilaton system in six and seven dimensions
for the elliptic, parabolic and hyperbolic cases.
As in lower dimensions \cite{Antonelli:2019dqv}, one can make use of various algebraic simplifications (e.g., the possibility of eliminating $u$ and its derivatives). Therefore, this new numerical procedure of obtaining scale-invariant solutions has significantly improved the actual computations with respect to the old setting \cite{Hirschmann_1997}, thus allowing to investigate more accurately $z_+$ crossing.

A new method of setting up the perturbation theory of self-similar solutions for elliptic and hyperbolic cases was recently proposed in \cite{Antonelli:2019dbi}. Given the self-similar solutions, one may perturb\footnote{Various perturbed solutions over spherical symmetric background for a specific theory had been revealed in \cite{Ghodsi_2010}.} the field $h(t,r)$ of the self-similar background solution letting
\begin{equation}
    h(t,r) = (-t)^{\Delta} \left( h_{\text{CSS}}(z) + \varepsilon \,(-t)^{-\kappa} h_1(z) \right)\,,
\end{equation}
with $\Delta$ the scaling dimension of the field $h$. One can also investigate solutions for the specific exponent $\kappa$ that finds all the modes. In \cite{Antonelli:2019dbi}, the Choptuik critical exponents $\gamma$ were found in four and five dimensions, relating them to the most relevant mode via the following equation \cite{KHA}:
\begin{equation}
   \frac{1}{\Re\kappa} = \gamma\,.
\end{equation}
We hope to be able to systematically study the perturbations for the generic parabolic case and also investigate the perturbations of the distinct solutions of this paper in the near future. It would be also interesting to investigate whether or not there could be elements $\Gamma\in \text{SL}(2,\mathbb{Z})$ such that
\begin{equation}
\tau(e^{\Delta_\Gamma}\,t,e^{\Delta_\Gamma}\, r)\,=\,{a\tau+b\over c\tau+d};\qquad \Gamma\,=\,\left(
\begin{array}{cc}a & b \\
c & d \end{array} \right)\in \text{SL}(2,\mathbb{Z})\ ,
\end{equation}
where $\Delta_{\Gamma}$ is the echoing parameter. In order to address this open question one need not assume CSS ad instead carry out the entire numerical integration of Einstein's equations. We hope to return to these open questions in the near future.
\section*{Acknowledgments}
We would like to thank R. Antonelli, E. Hirschmann, L. Alvarez-Gaume, I. Basile and A. Sagnotti for their useful comments. We are really grateful to R. Antonelli and E. Hirschmann for bringing to our attention various points on some of the numerical parts. EV carried out part of the work during her visit to Scuola Normale Superiore, and is grateful for the kind hospitality.
This work was supported by INFN (ISCSN4-GSS-PI), by Scuola Normale, and by the MIUR-PRIN contract 2017CC72MK\_003.

\end{document}